



%

%
 \font\twelvebf=cmbx12
 \font\twelvett=cmtt12
 \font\twelveit=cmti12
 \font\twelvesl=cmsl12
 \font\twelverm=cmr12		\font\ninerm=cmr9
 \font\twelvei=cmmi12		\font\ninei=cmmi9
 \font\twelvesy=cmsy10 at 12pt	\font\ninesy=cmsy9
 \skewchar\twelvei='177		\skewchar\ninei='177
 \skewchar\seveni='177	 	\skewchar\fivei='177
 \skewchar\twelvesy='60		\skewchar\ninesy='60
 \skewchar\sevensy='60		\skewchar\fivesy='60
%
%

%
 \font\fourteenrm=cmr12 scaled 1200
 \font\seventeenrm=cmr12 scaled 1440
 \font\fourteenbf=cmbx12 scaled 1200
 \font\seventeenbf=cmbx12 scaled 1440
%
%

%
%
%
\font\tenmsb=msbm10
\font\twelvemsb=msbm10 scaled 1200
\newfam\msbfam

%
\font\tensc=cmcsc10
\font\twelvesc=cmcsc10 scaled 1200
\newfam\scfam

%
\def\seventeenpt{\def\rm{\fam0\seventeenrm}%
 \textfont\bffam=\seventeenbf	\def\bf{\fam\bffam\seventeenbf}}
\def\fourteenpt{\def\rm{\fam0\fourteenrm}%
 \textfont\bffam=\fourteenbf	\def\bf{\fam\bffam\fourteenbf}}
\def\twelvept{\def\rm{\fam0\twelverm}%
 \textfont0=\twelverm	\scriptfont0=\ninerm	\scriptscriptfont0=\sevenrm
 \textfont1=\twelvei	\scriptfont1=\ninei	\scriptscriptfont1=\seveni
 \textfont2=\twelvesy	\scriptfont2=\ninesy	\scriptscriptfont2=\sevensy
 \textfont3=\tenex	\scriptfont3=\tenex	\scriptscriptfont3=\tenex
 \textfont\itfam=\twelveit	\def\it{\fam\itfam\twelveit}%
 \textfont\slfam=\twelvesl	\def\sl{\fam\slfam\twelvesl}%
 \textfont\ttfam=\twelvett	\def\tt{\fam\ttfam\twelvett}%
 \scriptfont\bffam=\tenbf 	\scriptscriptfont\bffam=\sevenbf
 \textfont\bffam=\twelvebf	\def\bf{\fam\bffam\twelvebf}%
 \textfont\scfam=\twelvesc	\def\sc{\fam\scfam\twelvesc}%
 \textfont\msbfam=\twelvemsb	
 \baselineskip 14pt%
 \abovedisplayskip 7pt plus 3pt minus 1pt%
 \belowdisplayskip 7pt plus 3pt minus 1pt%
 \abovedisplayshortskip 0pt plus 3pt%
 \belowdisplayshortskip 4pt plus 3pt minus 1pt%
 \parskip 3pt plus 1.5pt
 \setbox\strutbox=\hbox{\vrule height 10pt depth 4pt width 0pt}}
\def\tenpt{\def\rm{\fam0\tenrm}%
 \textfont0=\tenrm	\scriptfont0=\sevenrm	\scriptscriptfont0=\fiverm
 \textfont1=\teni	\scriptfont1=\seveni	\scriptscriptfont1=\fivei
 \textfont2=\tensy	\scriptfont2=\sevensy	\scriptscriptfont2=\fivesy
 \textfont3=\tenex	\scriptfont3=\tenex	\scriptscriptfont3=\tenex
 \textfont\itfam=\tenit		\def\it{\fam\itfam\tenit}%
 \textfont\slfam=\tensl		\def\sl{\fam\slfam\tensl}%
 \textfont\ttfam=\tentt		\def\tt{\fam\ttfam\tentt}%
 \scriptfont\bffam=\sevenbf 	\scriptscriptfont\bffam=\fivebf
 \textfont\bffam=\tenbf		\def\bf{\fam\bffam\tenbf}%
 \textfont\scfam=\tensc		\def\sc{\fam\scfam\tensc}%
 \textfont\msbfam=\tenmsb	
 \baselineskip 12pt%
 \abovedisplayskip 6pt plus 3pt minus 1pt%
 \belowdisplayskip 6pt plus 3pt minus 1pt%
 \abovedisplayshortskip 0pt plus 3pt%
 \belowdisplayshortskip 4pt plus 3pt minus 1pt%
 \parskip 2pt plus 1pt
 \setbox\strutbox=\hbox{\vrule height 8.5pt depth 3.5pt width 0pt}}

%
\def\twelvepoint{%
 \def\small{\tenpt\rm}%
 \def\normal{\twelvept\rm}%
 \def\big{\fourteenpt\rm}%
 \def\huge{\seventeenpt\rm}%
 \footline{\hss\twelverm\folio\hss}
 \normal}
%

%
\def\bigbold{\big\bf}

%
\catcode`\@=11
%
%
\def\footnote#1{\edef\@sf{\spacefactor\the\spacefactor}#1\@sf
 \insert\footins\bgroup\small
 \interlinepenalty100	\let\par=\endgraf
 \leftskip=0pt		\rightskip=0pt
 \splittopskip=10pt plus 1pt minus 1pt	\floatingpenalty=20000
 \smallskip\item{#1}\bgroup\strut\aftergroup\@foot\let\next}
%
%
%
%
\def\hexnumber@#1{\ifcase#1 0\or 1\or 2\or 3\or 4\or 5\or 6\or 7\or 8\or
 9\or A\or B\or C\or D\or E\or F\fi}
\edef\msbfam@{\hexnumber@\msbfam}

%
%
%
\catcode`\@=12

\newcount\EQNO      \EQNO=0
\newcount\FIGNO     \FIGNO=0
\newcount\REFNO     \REFNO=0
\newcount\SECNO     \SECNO=0
\newcount\SUBSECNO  \SUBSECNO=0
\newcount\FOOTNO    \FOOTNO=0
\newbox\FIGBOX      \setbox\FIGBOX=\vbox{}
\newbox\REFBOX      \setbox\REFBOX=\vbox{}
\newbox\RefBoxOne   \setbox\RefBoxOne=\vbox{}
\expandafter\ifx\csname normal\endcsname\relax\def\normal{\null}\fi

\def\Eqno{\global\advance\EQNO by 1 \eqno(\the\EQNO)%
    \gdef\label##1{\xdef##1{\nobreak(\the\EQNO)}}}
\def\Fig#1{\global\advance\FIGNO by 1 Figure~\the\FIGNO%
    \global\setbox\FIGBOX=\vbox{\unvcopy\FIGBOX
      \narrower\smallskip\item{\bf Figure \the\FIGNO~~}#1}}
\def\Ref#1{\global\advance\REFNO by 1 \nobreak[\the\REFNO]%
    \global\setbox\REFBOX=\vbox{\unvcopy\REFBOX\normal
      \smallskip\item{\the\REFNO .~}#1}%
    \gdef\label##1{\xdef##1{\nobreak[\the\REFNO]}}}
\def\Section#1{\SUBSECNO=0\advance\SECNO by 1
    \bigskip\leftline{\bf \the\SECNO .\ #1}\nobreak}
\def\Subsection#1{\advance\SUBSECNO by 1
    \medskip\leftline{\bf \ifcase\SUBSECNO\or
    a\or b\or c\or d\or e\or f\or g\or h\or i\or j\or k\or l\or m\or n\fi
    )\ #1}\nobreak}
\def\Footnote#1{\global\advance\FOOTNO by 1
    \footnote{\nobreak$\>\!{}^{\the\FOOTNO}\>\!$}{#1}}
\def\SameFootnote{$\>\!{}^{\the\FOOTNO}\>\!$}

\def\References{\bigskip\centerline{\bf REFERENCES}
                \smallskip\copy\REFBOX}
\def\NewRefPage{\setbox\RefBoxOne=\vbox{\unvcopy\REFBOX}
		\setbox\REFBOX=\vbox{}
		\def\References{\bigskip\centerline{\bf REFERENCES}
                		\nobreak\smallskip\nobreak\copy\RefBoxOne
				\vfill\eject
				\smallskip\copy\REFBOX}
		\def\NewRefPage{}}


\font\twelvebm=cmmib10 at 12pt
\font\tenbm=cmmib10
\font\ninei=cmmi9
\newfam\bmfam

\def\twelvepointbmit{
\textfont\bmfam=\twelvebm
\scriptfont\bmfam=\ninei
\scriptscriptfont\bmfam=\seveni
\def\bmit{\fam\bmfam\twelvebm}
}

\def\tenpointbmit{
\textfont\bmfam=\tenbm
\scriptfont\bmfam=\seveni
\scriptscriptfont\bmfam=\fivei
\def\bmit{\fam\bmfam\tenbm}
}

\tenpointbmit

\mathchardef\Gamma="7100
\mathchardef\Delta="7101
\mathchardef\Theta="7102
\mathchardef\Lambda="7103
\mathchardef\Xi="7104
\mathchardef\Pi="7105
\mathchardef\Sigma="7106
\mathchardef\Upsilon="7107
\mathchardef\Phi="7108
\mathchardef\Psi="7109
\mathchardef\Omega="710A
\mathchardef\alpha="710B
\mathchardef\beta="710C
\mathchardef\gamma="710D
\mathchardef\delta="710E
\mathchardef\epsilon="710F
\mathchardef\zeta="7110
\mathchardef\eta="7111
\mathchardef\theta="7112
\mathchardef\iota="7113
\mathchardef\kappa="7114
\mathchardef\lambda="7115
\mathchardef\mu="7116
\mathchardef\nu="7117
\mathchardef\xi="7118
\mathchardef\pi="7119
\mathchardef\rho="711A
\mathchardef\sigma="711B
\mathchardef\tau="711C
\mathchardef\upsilon="711D
\mathchardef\phi="711E
\mathchardef\cho="711F
\mathchardef\psi="7120
\mathchardef\omega="7121
\mathchardef\varepsilon="7122
\mathchardef\vartheta="7123
\mathchardef\varpi="7124
\mathchardef\varrho="7125
\mathchardef\varsigma="7126
\mathchardef\varphi="7127


\twelvepoint			

\font\twelvebm=cmmib10 at 12pt
\font\tenbm=cmmib10
\font\ninei=cmmi9
\newfam\bmfam

\def\twelvepointbmit{
\textfont\bmfam=\twelvebm
\scriptfont\bmfam=\ninei
\scriptscriptfont\bmfam=\seveni
\def\bmit{\fam\bmfam\twelvebm}
}

\def\tenpointbmit{
\textfont\bmfam=\tenbm
\scriptfont\bmfam=\seveni
\scriptscriptfont\bmfam=\fivei
\def\bmit{\fam\bmfam\tenbm}
}

\tenpointbmit

\mathchardef\Gamma="7100
\mathchardef\Delta="7101
\mathchardef\Theta="7102
\mathchardef\Lambda="7103
\mathchardef\Xi="7104
\mathchardef\Pi="7105
\mathchardef\Sigma="7106
\mathchardef\Upsilon="7107
\mathchardef\Phi="7108
\mathchardef\Psi="7109
\mathchardef\Omega="710A
\mathchardef\alpha="710B
\mathchardef\beta="710C
\mathchardef\gamma="710D
\mathchardef\delta="710E
\mathchardef\epsilon="710F
\mathchardef\zeta="7110
\mathchardef\eta="7111
\mathchardef\theta="7112
\mathchardef\iota="7113
\mathchardef\kappa="7114
\mathchardef\lambda="7115
\mathchardef\mu="7116
\mathchardef\nu="7117
\mathchardef\xi="7118
\mathchardef\pi="7119
\mathchardef\rho="711A
\mathchardef\sigma="711B
\mathchardef\tau="711C
\mathchardef\upsilon="711D
\mathchardef\phi="711E
\mathchardef\cho="711F
\mathchardef\psi="7120
\mathchardef\omega="7121
\mathchardef\varepsilon="7122
\mathchardef\vartheta="7123
\mathchardef\varpi="7124
\mathchardef\varrho="7125
\mathchardef\varsigma="7126
\mathchardef\varphi="7127



\relax
%


\twelvepointbmit		

\def\sqr#1#2{{\vbox{\hrule height.#2pt
 	\hbox{\vrule width.#2pt height#1pt \kern#1pt\vrule width.#2pt}
		\hrule height.#2pt}}}

\def\Partial#1{\partial_{#1}^{\raise2pt\hbox{$\scriptstyle 2$}}}
\def\bar{\overline}

\def\Uin{u^{\,\raise2pt\hbox{$\scriptstyle\rm in$}}}
\def\Uout{u^{\,\raise2pt\hbox{$\scriptstyle\rm out$}}}


\nopagenumbers

\def\today{\number\day\space\ifcase\month\or
  January\or February\or March\or April\or May\or June\or
  July\or August\or September\or October\or November\or December\fi
  \space\number\year}
\bigskip\bigskip

\null\bigskip\bigskip\bigskip

\baselineskip=27pt

\vskip 1cm

\centerline{\bigbold A SPINOR MODEL}
\centerline{\bigbold  FOR }
\centerline{\bigbold QUANTUM COSMOLOGY}
\bigskip\bigskip\bigskip

\centerline{\bf T Dereli  }
\centerline{\it Department of Mathematics,
Middle East Technical University}
\centerline{\it 06531 Ankara, Turkey}
\centerline{\tt dereli{$@$}trmetu.bitnet}

\medskip
\centerline{\bf M \"Onder }
\centerline{\it Department of Physics Engineering,
Hacettepe University }
\centerline{\it 06532 Ankara, Turkey}
\centerline{\tt F\_Onder{@}trhun.bitnet}
\medskip
\centerline{\bf Robin W Tucker}\medskip
\centerline{\it School of Physics and Materials,
University of Lancaster}
\centerline{\it Bailrigg, Lancs. LA1 4YB, UK}
\centerline{\tt rwt{@}lavu.physics.lancaster.ac.uk}

$$  \quad $$
\vfill



\eject

\vskip 2cm

\bigskip\bigskip\bigskip\bigskip

\centerline{\bf ABSTRACT}
\vskip 1cm

\midinsert
\narrower\narrower\noindent


The question of the interpretation of Wheeler-DeWitt solutions in
the context of cosmological models is addressed by
implementing the Hamiltonian constraint as a spinor wave equation in
minisuperspace.
We offer  a relative probability interpretation based on a
non-closed vector current in this space and a prescription for a
parametrisation of classical solutions in terms of classical time. Such a
prescription can accommodate classically degenerate metrics describing
manifolds with signature change. The relative probability density,
defined in terms of a Killing vector of the Dewitt metric on minisuperspace,
should permit  one to identify  classical loci corresponding to geometries
for a classical manifold. This interpretation is illustrated in the context
of a quantum cosmology model for two-dimensional dilaton gravity.

\endinsert

\vfill\eject


\headline={\hss\rm -~\folio~- \hss}     

\def\pprime{^\prime}
\def\frac#1#2{{#1\over #2}}

\Section{Introduction}

One of the many outstanding problems in trying to construct a quantum field
theory of gravitation concerns  the appropriate interpretation  of
quantum states for configurations that make no overt reference to ``time''.
Thus it is difficult in general to endow the theory with any traditional
Hilbert space structure based on a hermitian inner product and a unitary
evolution. Although many
alternative schemes have been suggested   difficulties in interpretation
remain. Some of the difficulties are intrinsic to the infinite dimensional
aspect of field quantisation and in this respect one often seeks guidance
by studying truncated field configurations corresponding to situations with
high symmetry. We shall not rehearse here the many cogent arguments that
urge caution in extending deductions from such models to the full quantum
field theory \Ref{Time and the Interpretation of Quantum Gravity, K V
Kuchar, Proceedings of the 4th Canadian Conference in General
Relativity and Relativistic Astrophysics, Eds. G Kunstatter, D Vincent, J
Williams, World Scientific (Singapore) 1992}. However symmetric models
 are a useful theoretical laboratory for testing
ideas that may have more general validity, and enable one to disentangle
conceptual problems from technical ones.
In the context of minisuperspace
models a number of authors have noticed that  it is
possible to implement the Hamiltonian constraint for Bianchi-type models in
 general relativity,
on a multicomponent wavefunction
\Ref{M P Ryan, Hamiltonian Cosmology (Berlin: Springer) (1972)}.
In an attempt to relate such states
to modes of the gravitino field such models have been examined in the
context of N=1 supergravity \Ref{ A Mac\'ias, O Obreg\'on M P Ryan, Class.
Quantum Grav. {\bf 4} (1987) 1477}  although the relation with the original
Wheeler-Dewitt equation is then lost \Ref{ P D D'eath, S W Hawking, Phys.
Letts. {\bf B300} (1993) 44}.
 In this letter we focus on a particular minisuperspace analysis
that gives rise to a Hamiltonian constraint, classically describing the
zero energy configuration of an oscillator ghost-oscillator pair. This
gives rise to a Wheeler-DeWitt equation that has occurred in a number of
different contexts.
 It appears in certain 4-dimensional spacetime
cosmologies
\Ref{L J Garay, J J Halliwell, G A Marug\'an, Phys. Rev.{\bf D43} (1991) 2572},
\Ref{ C Kiefer, Nucl. Phys. {\bf B341} (1990) 273},
\Ref{T Dereli, R W Tucker, Class. Quantum Grav.
{\bf 10} (1993) 365},
 \Ref{ T Dereli, M \"Onder, R W Tucker, Class. Quantum Grav. {\bf 10}  (1993)
1425},
 and we have discussed it in the context of a class of
2-dimensional dilaton-gravity models.
These models  have arisen
either from string-inspired limits or from the suppression  of inhomogeneous
modes in Einstein's theory of general relativity,
\Ref{G Mandal, A M Sengupta, S R Wadia, Mod.Phys. Lett. {\bf A6}  (1991)
1685},
\Ref{E Witten, Phys. Rev. {\bf D44}  (1991) 314 }\label\witten,
\Ref{C G Callan, Jr., S B Giddings, J A Harvey, A Strominger, Phys. Rev. {\bf
D45}  (1992) R1005 },
\Ref{S Chaudhuri, D Minic, ``On the Black Hole Background of
Two-Dimensional String Theory'', Austin Preprint UTTG-30-92 (1992) },
 \Ref{J Navarro-Salas, M Navarro,V Aldaya,
`` Wave Functions of the Induced 2D-Gravity'',
 Valencia Preprint FTUV/93-3}\label \ald ,
\Ref{ R Jackiw, `` Gauge Theories for Gravity on a Line'',
({\it In Memoriam} M C Polivanov), MIT Preprint 1993}
\Ref{M \"Onder, R W Tucker, Phys. Letts. {\bf B311} (1993) 47\label\MORT}.
 Our interest with this class of models
stems from the properties of coherent state solutions to the corresponding
 Wheeler-DeWitt equation and their relation to classical solutions to
general relativity including those that change signature
\Ref{M \"Onder, R W Tucker, On the Relation between Classical and Quantum
Cosmology in a 2D Dilaton-Gravity Model, Class. Q. Grav. (To Appear)
}\label\dilaton.
 We show in this
letter that it is possible to implement the Hamiltonian constraint in the
cosmological sector as a first order wave equation for a multicomponent
state vector and to endow the space of
 spinor  solutions to such an equation with a Hilbert space structure.
We offer an interpretation of such states in terms of  relative
probabilities defined by a non-conserved current.
This
is possible since such solutions enable one to construct such a
 current with a positive definite density defined by the Killing
isometry of the minisuperspace metric.
We suggest that such an
interpretation is not unnatural in a quantum theory
 that attempts to accommodate states whose classical limits describe
manifolds  with degenerate metrics where the signature can change. Such
limits correspond to more exotic spaces
 in which the global topology may
be non-trivial and the geometry non-Riemannian.
In such situations  the emergence of an arrow of classical time may have its
origin in an underlying quantum description of such spaces.

\Section{The Model}

We have recently  developed {\dilaton} a canonical quantisation of the
2-dimensional dilaton-gravity theory based on the classical action
$$
S[g,\psi]=\int_N\,\,\left\{ \frac 12\psi\star{\cal R}+cd\psi\wedge\star
d\psi +\star (\Lambda_0+\alpha e^{c\psi})\right\}\Eqno$$
where $N$ is some domain of a two-dimensional manifold, $\psi$ is a real
scalar field and ${\cal R}$ is the curvature scalar of the Levi-Civita
connection associated with the metric tensor $g$. The operator $\star$
denotes the Hodge map of $g$ and $c,\Lambda_0$ and $\alpha$ are constants.
The classical cosmological sector of this theory can be solved exactly and
admits solutions with a degenerate metric where the signature changes from
being Lorentzian to Euclidean. The standard approach for implementing the
Hamiltonian constraint in the quantum version of such theories is to search
for complex scalar valued functions on the appropriate manifold of matter
and space geometry configurations. Thus in {\dilaton } we took $\bf {R^2}$ as a
minisuperspace with global coordinates $\{X,Y\}$ labeling these
configurations and sought Wheeler-DeWitt solutions $\Psi: \bf {R^2}\mapsto \bf
{C}$ to the
equation:
$$H\Psi=0\Eqno \label\wdwitt $$
where $$H=(\omega^2 X^2-\frac{\partial_X^2}{4})-
(\omega^2 Y^2-\frac{\partial_Y^2}{4})\Eqno$$
The Wheeler-DeWitt equation \wdwitt\ endows ${\bf R}^2$ with a natural
(Lorentzian signature) metric ${\cal G}$. In terms of the coordinates
$\{X,Y\}$:
$${\cal G}=\partial_X\otimes \partial_X -\partial_Y\otimes \partial_Y.\Eqno
\label\metric$$
If $\#\,$ denotes the associated Hodge map then \wdwitt\ may be written:
$$d\#\,d\Psi-W\#\,\Psi=0\Eqno \label\wwdw$$
where $W(X,Y)=4\omega^2(X^2-Y^2)$. By multiplying \wwdw\ by
$\bar\Psi$ and subtracting from the corresponding equation obtained by
complex conjugation we readily verify that
$$d\,{\cal J}=0\Eqno$$
where the current 1-form
$${\cal J}=Im(\bar\Psi \# d\,\Psi).\Eqno \label\kgcurrent$$
Although this current is conserved there is no preferred spacelike
foliation of ${\bf R^2}$ that defines a ``density'' component of
${\cal J}$ that does not in general change sign. Furthermore, although the
minisuperspace is flat there is no natural way to restrict solutions to have a
positive definite norm. The
Killing vectors of the metric {\metric } do not generate a symmetry of the
equation \wwdw. Thus there appears no invariant way to normalise solutions
of \wwdw, construct a Hilbert subspace of normalisable solutions and endow
the quantum theory with the standard probabilistic interpretation. By
contrast the traditional Klein-Gordon quantisation of the relativistic
free particle in Minkowski spacetime exists because the Killing isometry of
the  spacetime metric induces a classical symmetry of the Klein-Gordon
equation. Furthermore one can then exploit  translational symmetry to
restrict solutions to either the positive or negative mass-hyperboloid in
the space of spatial Fourier modes on which the above current induces a
positive-definite inner-product. However if one considers the quantisation
in a non-stationary spacetime (or in a stationary spacetime with a time
dependent potential) again one may loose the timelike Killing
symmetries that enable one to effect the above construction and the
particle interpretation of the field system is at best an asymptotic
notion in a second quantised formulation.
Before the advent of second quantisation Dirac was motivated to implement
the constraint arising from the reparametrisation of the action for a
relativistic free particle by a first order equation for the space of
quantum states. In a similar vein we are interested here in the possibility of
implementing
the constraint {\wdwitt } by a first order equation for a complex
multicomponent field $\Phi: {\bf R^2}\mapsto{\bf C^n}$ for some ${\bf n}$, such
that each component of $\Phi$ satisfies \wdwitt. If this is possible it is
natural to seek for a  current constructed from $\Phi$ that admits a
positive-definite charge density for some class of  foliations that are
spacelike with reference to the metric (4). If this is possible then the
choice of a probability interpretation is determined by the Lorentzian
structure of (mini-)superspace. (This Lorentzian structure has its origins
in the universal gravitational attraction between matter and must be clearly
distinguished from the light-cone structure of classical spacetime).
Such a choice  seems natural if we wish to extend the
interpretation of the theory to accommodate  classical limits that include
spacetime metrics with a  signature transition.
In a domain with Euclidean signature  one has no natural
means of defining the spacelike
and timelike components of a current.  Thus the definition of a probability
current must transcend any definition of any preferred
classical time for classical spacetimes. We shall reiterate our views on the
latter problem in the last section.

\Section{The Clifford Algebra of Minisuperspace}

\def\v{\vee}

The natural Lorentzian null-cone
structure in  2D minisuperspace
 endows the space with a (1,1) Clifford bundle structure
\Ref{ I M Benn, R W Tucker, An Application to Spinors and Geometry with
Applications in Physics, (Adam Hilger) (1987)}\label\book .
Thus there is a matrix basis for the Clifford Algebra
Cl(1,1) in which the 1-forms $dX$ and $dY$ are represented as matrices
satisfying
$$ dX \v dX=1\Eqno $$
$$ dY \v dY=-1\Eqno $$
$$ dX \v dY+dY\v dX=0\Eqno $$
where $\v$ denotes multiplication in the Clifford algebra.
In conventional gamma matrix notation:
 ($dX \mapsto \gamma^1,dY \mapsto \gamma^0$).
The Clifford bundle has minisuperspace as base and Cl(1,1) as fibre.
Let $\Phi$ be a section of this bundle:
$$\Phi=\Phi_0+\Phi_1 \,dX +\Phi_2 \,dY+\Phi_{12}\,dX \v dY.\Eqno $$
Since the bundle is trivial the components
${\Phi_j}\equiv \{\Phi_0,\Phi_1 ,\Phi_2 ,\Phi_{12}\}$ may be regarded as
complex functions
on $\bf R^2$.

\def\dslash{{\cal D}}

Introduce the Clifford potentials:
$$V_1=2i\omega (-Y+X \,dX \v dY)\Eqno $$\label\pota
$$V_2=2i\omega (Y+X \,dX \v dY).\Eqno $$\label\poyb
We assert that if $\Phi$ satisfies first order equation:
$$\dslash\Phi+V_1 \v \Phi=0 \Eqno \label\kdwitt $$
where $\dslash=(d-\delta)$  then each component of $\Phi$ will satisfy \wdwitt.
In this equation $d$ denotes exterior differentiation and $\delta$ is the
coderivative:
$$\delta=\#\,^{-1}d\,\#\,\eta$$
where the involution $\eta$ {\book} is a linear operator on $\Phi$ that
preserves
0-forms, reverses the sign of 1-forms and
$$\eta (dX \v d Y)=d X \v dY.\Eqno$$
The above result follows from
$$(\dslash +V_2)\v (\dslash +V_1)=(\dslash^2-W) \Eqno$$
and the recognition that $\dslash^2=-(d\,\delta+\delta\, d)$
is the Laplace-Beltrami
operator. Since the basis forms in $\Phi$ are all holonomic it follows that
if $\Phi$ satisfies {\kdwitt } then its components satisfy
$$d\#\,d\Phi_j-W\#\,\Phi_j=0.\Eqno$$

\Section{Spinor Solutions and Associated Currents}
\def\j{\cal I}

We observe that since our minisuperspace is flat with respect to {\metric} it
is possible to find solutions $\Psi$ of {\kdwitt } that lie in a minimal
(left) ideal of Cl(1,1) at each point. For example we may decompose
$$\Phi=\Phi \v P_{+}+\Phi \v P_{-}\Eqno$$
where $P_{\pm}=\frac 12 (1\pm d X)$ and take $\Psi=\Phi\v P_{+}$.
Minimal ideals provide irreducible modules for the  sub-group SPIN of the
Clifford group of Cl(1,1) {\book}
and their elements are spinors. Since $P_{+}$ is a parallel idempotent in
minisuperspace, if $\Phi$ is a solution of {\kdwitt } then so is
$\Phi\v P_{+}$ so
$\Psi$ may be regarded as a spinor solution of {\kdwitt }.

We concentrate on those  spinor solutions of (14) that are
asymptotically well behaved as $\vert X\vert $ or $\vert Y\vert $
 tends to infinity.
Such  solutions may be expressed in terms of a basis of Hermite functions.
Thus if $$ \Psi=(u - v dY)\v P_{+}  \label\spinorsol \Eqno $$
(14) may be expressed as the coupled partial differential equations:
$$ \partial_\xi F-2i\omega \xi \,H=0 \Eqno$$
$$ \partial_\eta H+ 2i\omega \eta \,F=0 \Eqno$$
where $\xi=(Y-X)/\sqrt(2)\,\,\eta=(Y+X)/\sqrt(2),\,\,F\,=u-v,\,\, H=u+v$.
These have the solutions
 $$F(X,Y)=\sum_{n=0}^{\infty} c_n e^{-(z_1^2+z_2^2)/2} H_n(z_1)H_n(z_2)\Eqno $$
 $$H(X,Y)=\sum_{n=0}^{\infty} b_n e^{-(z_1^2+z_2^2)/2} H_n(z_1)H_n(z_2)\Eqno $$
where $z_1=\sqrt(2\omega)X$ and $z_2=\sqrt(2\omega)Y$.
 The complex
coefficients
$\{c_n\}$ and  $\{b_n\}$ are linearly correlated by the wave equation (14). A
typical `` coherent spinor state `` solution takes the form:
$$F=C e^{-(\alpha(X^2+Y^2)-2\beta X Y)} \Eqno$$
$$H=\frac{i}{\omega}(\alpha+\beta)F\Eqno$$
where $\alpha , \beta , C$ are arbitrary complex constants.
Since our 2-dimensional minisuperspace is topologically trivial it is
always possible to find closed 1-forms that are candidates for conserved
currents. However it is non-trivial to construct  a current that has a
positive definite density for a class of solutions to \kdwitt.
It is not difficult to verify that the (complex) current
$${\cal J}\pprime = H^2\xi\, d\xi-F^2\eta\,d \eta\Eqno $$
is closed and hence gives rise to a conserved current in mini-superspace.
However for general solutions of (14) there is no  foliation
of mini-superspace that enables one to construct a non-negative real density
from such a conserved current.

We recall that in the Dirac theory of a relativistic particle described by
a spinor $\psi$ on spacetime,
the Dirac vector current with components $\bar\psi
\gamma^\mu \psi$ is conserved and possesses a positive-definite
density for any non-trivial spinor. In the language of Clifford bundles
this current is the form:
$$j[\psi]=*Re S_1(\psi\v \tilde\psi)\Eqno \label\current$$
where $S_1$ projects out the 1-form part of its argument and
$\tilde\psi=C^{-1}\v \psi^{\j}$ for some involution $\j$ in the (simple)
Clifford algebra that is equivalent to hermitian conjugation:
$$A^\dagger =C^{-1} A^{\j} C\Eqno$$
for all elements $A$ in the Clifford algebra.

With this goal in mind
we find from {\kdwitt} and {\current}
$$dj[\Psi]=Re\,Tr(\tilde\Psi\v (V_2-V_1)\v \Psi)\v \#1\Eqno\label\diveq \, $$
for all spinor solutions $\Psi$ of \kdwitt $\,\,\,$ where
$$j[\Psi]=\# Re S_1(\Psi\v \tilde\Psi).\Eqno \label\current$$
Here transposition is induced by the involution ${\cal I} \equiv\eta\xi$ where
$\xi$ is the main
anti-involution {\book} of the Clifford algebra and the adjoint spinor
$\tilde\Psi=C^{-1}\vee \bar{\Psi}^{\xi\eta }$, where $\bar{\Psi}$ denotes
the complex conjugation of $\Psi$. In the spinor basis defined by the
projectors $P_{\pm}$
that we are  using, the element $C=dY$.
It follows from
{\diveq} that $j[\Psi]$ defines a closed current for solutions that satisfy
the condition $Re\,Tr(i \omega\tilde\Psi\v \Psi)=0$. Although such
solutions do exist we shall not impose this restrictive condition in the
following discussion.
\Section{Discussion}

For a spinor solution (19)
\spinorsol
$\,\,\,$  with components $\{u,v\}$ we find in the
chart $\{X,Y\}$:
$$j[\Psi]=uv\,d Y-(\frac 12 \vert u \vert^2+\frac 12 \vert v \vert^2)\, d
X\Eqno$$
which clearly displays the non-negativity of the density
$$\rho_{K}(X,Y)=-j[\Psi](K)\Eqno$$
where $K$ denotes the Killing vector field ${\partial_X}$.
If $\Psi$ carries a representation of SPIN, then such a density, defined by
a spacelike Killing vector field will remain
positive for all proper Lorentz transformations that preserve the metric.
Thus it may be adopted as a probability measure for the interpretation of
the theory. However as the notation indicates a choice of spacelike
 Killing vector $K$ is implied. Since the current $j[\Psi]$ is not
conserved for all $\Psi$
there exists no choice of spacelike foliation such that the integral of the
probability density over a particular leaf of the foliation
is independent of the leaf chosen.
The existence of such a leaf dependent ``charge''
means that one cannot identify such a leaf as an ``instant of time'' and
interpret such a charge as a normalisation factor for a
 state in the traditional manner.
If we adopt as
the Hilbert space  norm of a
state $\Psi$
$$(\Psi,\Psi)_{K}=\int_{\bf R^2} \rho_{K} \# 1\Eqno$$
then this also will depend on the choice of Killing vector $K$.
Inasmuch as any convenient norm can be used to define  a Hilbert space this
is not necessarily a drawback. However the probabilistic interpretation of the
theory must be restricted to describing relative probabilities between
configurations. Thus for $K={\partial_X}$ the relative probability densities
for  configuration $Y_1 $ and $Y_2 $
irrespective of X may be defined as $\mu(Y_1)/\mu(Y_2)$ where:
 $$\mu(Y)=\int_{-\infty}^{\infty} \rho_K(X,Y) \,d X\Eqno$$
and in general $\mu(Y)$  will depend on the configuration
variable $Y$.

Up to this point no mention has been made of classical time. Indeed until
this issue is resolved there is little to recommend any particular
probabilistic interpretation of the theory since the obvious questions that
need to be addressed involve classical observers in a classical spacetime.
Thus we assert that in order to give substance to the above one should
concentrate on particular
quantum states that  can be related to classical
cosmologies.
As we have stressed we do not wish to exclude from such classical cosmologies
those that admit degenerate spacetime metrics. Thus we focus on those
solutions that for a given choice of Killing vector $K$ enable one to
constructs functions $\rho_K$ that have maxima in the vicinity of those
loci  in (mini-)superspace corresponding to parametrised
solutions to the classical field equations. For a classical manifold
with a proscribed topology and a proscribed signature structure we
concentrate on
particular classical solutions with degenerate metrics. Furthermore the
manifold should enable one to perform a Hamiltonian description of the
field equations so that the classical and quantum degrees of freedom can be
put into correspondence {\dilaton}.
In the cosmological context of the model in this discussion such a
correspondence is given by  a parametrised curve in mini-superspace:
$$\tau \mapsto
\{X=X(\tau),Y=Y(\tau)\}\quad\quad\quad \tau_0 < \tau <\tau_1\Eqno$$
Such a parametrisation of a classical solution may describe a Euclidean
signature metric  for part of the manifold and a Lorentzian signature
metric elsewhere. Thus it is natural to use  $\tau$ as a choice of
classical evolution parameter which is a classical time in the Lorentzian
domain. We may now transfer the relative probability
interpretation to the class of classical observers that inhabit the
classical cosmology defined by the locus of the maxima of $\rho_K$.
The density $\rho_K(X(\tau),Y(\tau))$ now offers a means of predicting the
relative probabilities for finding the classical configurations $\{X,Y\}$
at ``times'' $\tau_1$ and  $\tau_2$. The freedom in choosing  different
parametrisations to describe the same classical solutions corresponds to
the freedom in choosing different coordinate systems on the classical
manifolds.

   To illustrate the viability of this approach within the context of the
model defined by (2) we have sought a Killing vector $K$ that enables one to
construct a ``coherent state'' that can be used to construct a classical
spacetime in the vicinity of its peak.
Define $ \Psi_{s_1,s_2}=(u - v dY)\v P_{+}$
with
        $$u=s_1 e^{c_1(X^2+Y^2)+2c_2 \,XY}\Eqno$$
        $$v=s_2 e^{c_1(X^2+Y^2)+2c_2 \,XY}\Eqno$$
for some complex constants $c_1,c_2,s_1,s_2$. This is a solution to (14)
provided $$c_1=\frac{2i\omega s_1 s_2}{s_1^2-s_2^2}\Eqno$$
         $$c_2=\frac{i\omega(s_2^2+s_1^2)}{s_2^2-s_2^2}.\Eqno$$
Then for suitable $\{s_1,s_2,s_3,s_4\}$ the superposition
$$\Phi=\Psi_{s_1,s_2}-\Psi_{s_3,s_4 }\Eqno$$
enables one to construct a density $\rho_{\partial_X}$ that peaks along
classical loci for the theory defined by the action (1).
This is illustrated
in Fig 1 for the choice
$\{s_1=(3+0.1i),\,s_2=(1.3+0.1i),\,s_3=(3.1+0.1i),\,s_4=(1.4+0.1i)\}$.
The classical solutions correspond to the elliptical contours defined by
this density profile and have been discussed in {\dilaton}.
A notable feature of this state is the existence  of a particular locus among
the classical solutions for which the
divergence of the vector current $j[\Phi]$ does in fact vanish. In this
sense one may say that there is approximate conservation in the vicinity of
this particular classical configuration

Equation {\diveq } is reminiscent of the equation that follows from the
non-relativistic \break
Schr\"odinger equation in the presence of a complex
potential. Indeed the lack of hermiticity of the hamiltonian there is analogous
to the
property $ V_1\neq \pm V_2^{\cal I}$. In the Schr\"odinger situation the
use of a complex potential models the absorptive properties of an open
system. For a closed system  a non-hermitian hamiltonian is usually
regarded as pathological. However in the context of gravitation such a
reaction requires caution \MORT. For example
if the non-unitary evolution of a pure
state of matter to a mixed state via the Hawking process can be maintained
when gravitational back reaction is taken into account then probability
conservation in a gravitational context may not be tenable. It is clear
from the behaviour of the state in Figure 1 why the  conservation of our
current is impossible. Since the state vanishes asymptotically in all
directions in the configuration space there is no way that a
flux of positive density from the peaks of the state can flow smoothly to zero
{}.
In such a scenario it is tempting to conjecture that it is the existence of
 degenerate classical geometries that are mandatory to accommodate the
absorption of probability flux in the Euclidean domains. Just as the
creation (and annihilation) of a classical
cosmology may correspond to such  domains where a classical spacetime
description breaks down, the same may be true at the end points of localised
gravitational collapse. Of course a cosmological model is insensitive to
the subtleties required to accomodate a full quantisation of such a system.
However it would be a novel approach to implement the untruncated
canonical constraints
in terms of a first order  set of functional differential equations for a
multicomponent state vector such that each component
satisfies the traditional Wheeler-DeWitt equation.

\Section{Conclusion}

For our particular model we have chosen a symmetry vector of the
DeWitt metric on superspace that enables one to construct, from a
particular  quantum
state, a density that has maxima in the vicinity of classical cosmological
loci. An
internally consistent interpretation for such a density is provided in terms
of relative probabilities of the occurrence of classical matter and
a cosmological metric.  In general the associated current is not closed on
superspace  although the divergence
 is zero in the vicinity of certain classical cosmologies.
Whether the use of a non-conserved probability current has other
implications for quantum cosmology will be pursued elsewhere.

\Section{Acknowledgments}

This research was supported in part by the European Union under the Human
Capital and Mobility programme.
The authors are also grateful to NATO Research Grant CRG 891009.
T D and M \" O acknowledge partial support by the Scientific and Technical
Council of Turkey (TUBITAK) under TBAG-\c{C}G-1

\vfill\eject

\References

\vfill\eject

\vskip 2cm

{\centerline {Figure 1}}

\vskip 12cm

$\vert \Psi_{s_1,s_2}-\Psi_{s_3,s_4 }\vert ^2$
for the choice
$\{s_1=(3+0.1i),\,s_2=(1.3+0.1i),\,s_3=(3.1+0.1i),\,s_4=(1.4+0.1i)\}$.
The peak accentuates a
classical solution for the theory defined by the action (1).

\bye


\bye